\documentclass[useAMS,usenatbib]{mnras}

\usepackage{graphicx}
\usepackage{url}
\usepackage{multicol}

\title[Brown-Dwarf Ages]{Globular cluster absolute ages from cooling brown dwarfs}

\author[Caiazzo et al.]{Ilaria Caiazzo\thanks{E-mail:  ilariacaiazzo@phas.ubc.ca}$^{1}$, Jeremy S. Heyl$^{1}$, Harvey Richer$^{1}$, Jason Kalirai$^{2}$\\
\\
$^{1}$ Department of Physics and Astronomy, University of British Columbia, 6224 Agricultural Road, Vancouver, BC V6T 1Z1, Canada\\
$^{2}$ Space Telescope Science Institute, Baltimore MD 21218 
}

\begin{document} 

\date{Accepted ---. Received ---; in original form ---}

\pagerange{\pageref{firstpage}--\pageref{lastpage}} \pubyear{2017}

\maketitle 

\label{firstpage}

\begin{abstract}
Globular clusters are the oldest conglomerates of stars in our Galaxy and can be useful laboratories to test theories from stellar evolution to cosmology. In this paper, we present a new method to estimate the absolute age of a globular cluster from observations of its brown dwarfs. The transition region between the end of the main sequence and the brown dwarf regime is characterized by a dearth of objects as function of magnitude. The brightest of the cooling brown dwarfs is easily identified by an increase in density in the color magnitude diagram as you go fainter in magnitudes, and these brightest brown dwarfs get fainter with age. By identifying the brightest brown dwarfs, it is thus possible to determine the age of a globular cluster within a 1 Gyr precision with four-sigma confidence. This new method, which is independent of current methods of age estimation and which does not rely on the knowledge of the cluster's distance from Earth, will become feasible thanks to the high spatial resolution and incredible infrared sensitivity of the James Webb Space Telescope. 
\end{abstract}


\section{Introduction}
\label{sec:introduction}

Globular clusters are approximately spherical collections of stars, held together by their self-gravity, that orbit the center of their host galaxies and which can be found both in the halo of the galaxy and in the disk.  Globular clusters are particularly interesting objects as their observation can provide useful information for both stellar astrophysics and cosmology. Stars in a globular cluster comprise an approximately uniform population in age, chemical composition and distance from Earth. For this reason, globular clusters are ideal laboratories for testing stellar evolution models. On the cosmological side, globular clusters are the oldest objects in the Milky Way Galaxy for which an age can be determined and thus, their absolute ages can provide a lower limit to the age of the Universe. Measuring the age of Galactic globular clusters is also a crucial step in understanding the formation and evolution of our Galaxy. Determining the relative age between metal-rich clusters, found mainly in the inner Galaxy, and metal-poor clusters, located in the halo, can tell us which were formed inside the Galaxy and which are more likely to have been formed in dwarf galaxies, later swallowed by the Milky Way, providing clues to the Galaxy's formation. 

Several methods have been used to determine the age of globular clusters. The most straightforward way is to compare the observed properties of turn-off and giant-branch stars with stellar evolution models \citep{2005AJ....130..116D,1999AJ....118.2306R,2011ApJ...738...74D}. As stars evolve, their luminosity and surface temperature change, so that at each stage of their evolution, they occupy a specific position on the Hertzsprung-Russell (temperature-luminosity or HR) diagram. Stellar evolution models can predict a star's luminosity and effective temperature at each stage of its life, so that by comparing the observed properties of stars in a cluster with the model predictions, we can estimate the cluster's age. Current models are better constrained for main sequence stars, so that the time it takes for a star to consume all the hydrogen in its core is one of their most solid predictions. Stars that have reached this stage leave the main sequence and move in the HR diagram toward the red-giant branch, creating a well distinguishable feature called the turn-off.  The turn-off method provides an estimate of the age of the cluster by comparing this feature and the observed properties of red giant branch stars with stellar-evolution model predictions. This method, however, is subject to some uncertainties. On one hand, stellar evolution models depend on a range of physical parameters, which are still affected by theoretical uncertainties. On the other hand, the observed properties of stars in globular clusters do not depend only on age, but also on the stars' metal abundance and on the distance of the cluster from Earth.  It is difficult to decouple the different correlations. It is also worth mentioning that what we observe from Earth is not the stellar luminosity and temperature, but their magnitudes in different filters. In order to convert these observations into parameters that we can compare with our stellar evolution models, we have to employ stellar atmosphere models, which present theoretical uncertainties of their own. 

The turn-off method has been the most used in the literature to estimate the age of globular clusters. However, the need for other methods does not come only from its theoretical uncertainties,  but also from the fact that its sensitivity to age is quite poor, especially for old clusters. In old globular clusters, the turn-off stars have low masses, less than a solar mass, and low mass stars do not change appreciably in luminosity and temperature as they age; it is thus hard to use the turn-off feature to make an age estimate with a precision better than a few Gyr. There are, however, other features in the HR diagram that are more sensitive to age and that can be compared to stellar evolution models to estimate the age of clusters.  One feature that is sensitive to age, and which has already been employed to estimate the age of globular clusters \citep{2013Natur.500...51H},  is the white dwarf cooling sequence. White dwarfs are the last stage in the evolution of stars less massive than approximately eight times the mass of the Sun. They have no nuclear burning in their core, so their structure is supported by electron degeneracy pressure. Because there is no energy production in their core and yet they emit energy into space as electromagnetic radiation (and neutrino radiation when they are very young), they cool down and get fainter with age. By comparing a cluster's observed population of white dwarfs with the one expected from cooling models, it is possible to get an estimate of the cluster's age: the luminosity function's peak gets fainter with age. However, uncertainties in the white-dwarf equation of state, composition and atmospheres can affect these results.

In this paper we present a new, independent way of determining the age of globular clusters from the observation of cooling brown dwarfs. Brown dwarfs (BDs) are interesting objects per se as they represent a bridge between stars and planets. However, their formation and evolution are poorly understood, mainly because it is hard to identify them due to their faintness. Although it is possible to observe young and bright brown dwarfs in the field  (Two Micron All Sky Survey,\citealt{2006AJ....131.1163S,1999AJ....118..997G}, the Sloan Digital Sky Survey
\citealt{2000AJ....120.1579Y,2002AJ....123.3409H}, the United Kingdom Infrared Telescope Deep Sky Survey \citealt{2007MNRAS.375..213W,2007MNRAS.374..372L,2010MNRAS.406.1885B}, and the Wide-field Infrared Survey Explorer \citealt{2010AJ....140.1868W,2011ApJS..197...19K}), if the BDs are not in clusters, properties like distance and chemical composition, are difficult to determine.  On the other hand, with observations of brown dwarfs in globular clusters we could determine these properties with greater confidence. In Galactic globular clusters brown dwarfs have been cooling since nearly the beginning of time, getting fainter and redder.  At present, in most globular clusters, BDs are too faint to be observed. A recent attempt has been made to identify brown dwarfs in the globular cluster M4 \citep{2013Natur.500...51H}. The James Webb Space Telescope (JWST), which will be launched in October 2018, will provide observations of globular clusters much deeper than currently active telescopes, thanks to its sensitivity, especially in the infrared bands, where the radiation from the $\sim 1,000~\mathrm{K}$ brown dwarfs peaks.

Brown dwarfs are sub-stellar objects that failed to ignite or sustain hydrogen burning. Using the Modules for Experiments in Stellar Astrophysics (MESA) stellar evolution code~\citep{2011ApJS..192....3P}, we generated stellar evolution models, with a focus on the transition between low mass stars and BDs. From the models, we built isochrones,  {\em i.e.} collections of stars with a range of masses that simulate a globular cluster at a certain age.  The isochrones show that although there is a small difference in mass between low-mass stars and brown dwarfs, there is a large difference in luminosity; therefore, we expect few objects to lie in this large region of the HR diagram, and so the density of stars in the diagram drops significantly.  In order to calculate how this feature would look in a color-magnitude diagram (CMD) observed with JWST, we employed the stellar atmospheres developed by \citet{2015A&A...577A..42B} and \citet{2016Allard}. The resulting CMDs in the JWST filters show that the fainter end of the main sequence is followed by a gap of an order of  magnitude in luminosity and a large gap in color, in which the number of stars is much lower than on the main sequence.  At the end of the gap, where we find the cooling BDs, the density of objects increases again, presenting an easily detectable benchmark. Since BDs are cooling, the position in magnitude of the brightest BDs gives a clear indication of the age of the cluster.

\section{Mind the gap}
\label{sec:models}

In order to probe the evolution of brown dwarfs, we ran a suite of models \citep{2011ApJS..192....3P} for different initial masses and the metal content of the globular cluster 47~Tucanae. The specifications of our models are explained in Appendix~\ref{sec:mesamodels}. From the models, we can see that in the early stages of their evolution,  BDs evolve exactly like pre-main sequence stars, and this pre-main sequence phase, in which they contract and become fainter and bluer, can last for billions of years, depending on the mass of the object.

\begin{figure}
  \includegraphics[width=\columnwidth,clip,trim=20 5 45 20]{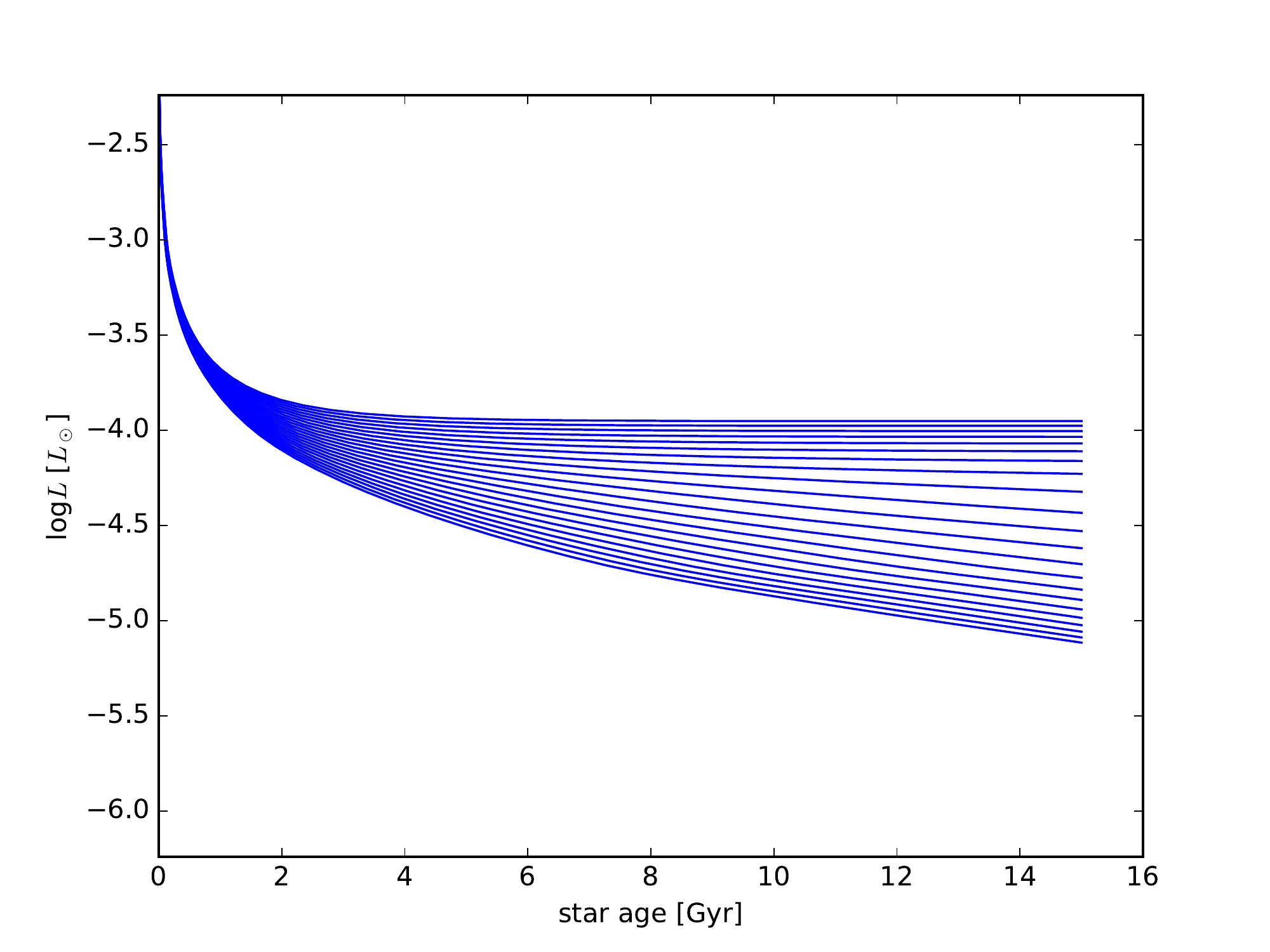}
  \caption{The evolution in luminosity vs age is plotted for 22 stars evenly spaced in mass between 0.0679 and 0.0763 M$_\odot$, from bottom to top. The evolution tracks come from our MESA models (see Appendix~\ref{sec:mesamodels}). Notice the increase in the density of the tracks at the top and bottom relative to the central region and how the lower tracks grow ever fainter.  The low density central region, which widens with age, corresponds to the gap between low mass main sequence stars and cooling brown dwarfs.  }\label{fig:lmod}
\end{figure}

Fig.~\ref{fig:lmod} depicts the evolution in luminosity for stars in this transition region between low mass stars and BDs. The mass range is for illustration only, since, for these models, we used the \texttt{tau\_1\_tables} atmosphere boundary conditions, which are available for solar metallicity only (see Appendix~\ref{sec:mesamodels}). For a lower metallicity cluster, like 47~Tucanae, the transition mass is closer to 0.08 M$_\odot$. This is due to the fact that a higher metallicity atmosphere is characterized by a higher opacity. Stars in this phase are fully convective, so a higher opacity in the atmosphere allows the star to retain more heat, favoring the ignition and the sustenance of nuclear burning in the core.

From the figure, we can see how the evolutionary path changes in the transition between low mass stars and brown dwarfs. Low mass stars get fainter with time in the pre-main sequence phase, where the bulk of their luminosity is provided by gravitational contraction and the temperature in their core is not high enough for nuclear burning to proceed at a high pace. For these low-mass stars, this phase lasts for up to 6 Gyr, depending on the star's mass, after which nuclear burning gets faster in the stars' core and becomes the main contribution to the production of energy, stabilizing the stars against gravitational contraction. In the plot, this moment is characterized by an increase in luminosity and represents the beginning of the stars' main sequence.  In brown dwarfs, on the other hand, temperatures in the core never get high enough to stabilize nuclear burning. Burning in their core decreases with time and they keep cooling at a steady rate. This difference in behavior creates a gap between the two types, in which a handful of intermediate masses keep cooling but at a slower rate, due to the presence of some hydrogen burning in their core. This behavior was earlier predicted in the brown dwarf evolution models by \citet{1989ApJ...345..939B}.

Note that the density in luminosity doubles at the end of the gap, {\em i.e.} at the top of the brown dwarf cooling sequence, which in the plot corresponds to a mass of 0.070 M$_\odot$.  For a cluster with a lower metal content such as 47 Tucanae, the top of the cooling sequence lies at a mass of 0.079 M$_\odot$. The size of the increase in density does not change with time, but the gap gets larger with age as the higher mass brown dwarfs cool, providing a clear indication of the age of the cluster.  This is independent of the current methods of age estimation.

\section{Observing brown dwarfs with JWST}
\label{sec:JWST}

\begin{figure}
\includegraphics[width=\columnwidth]{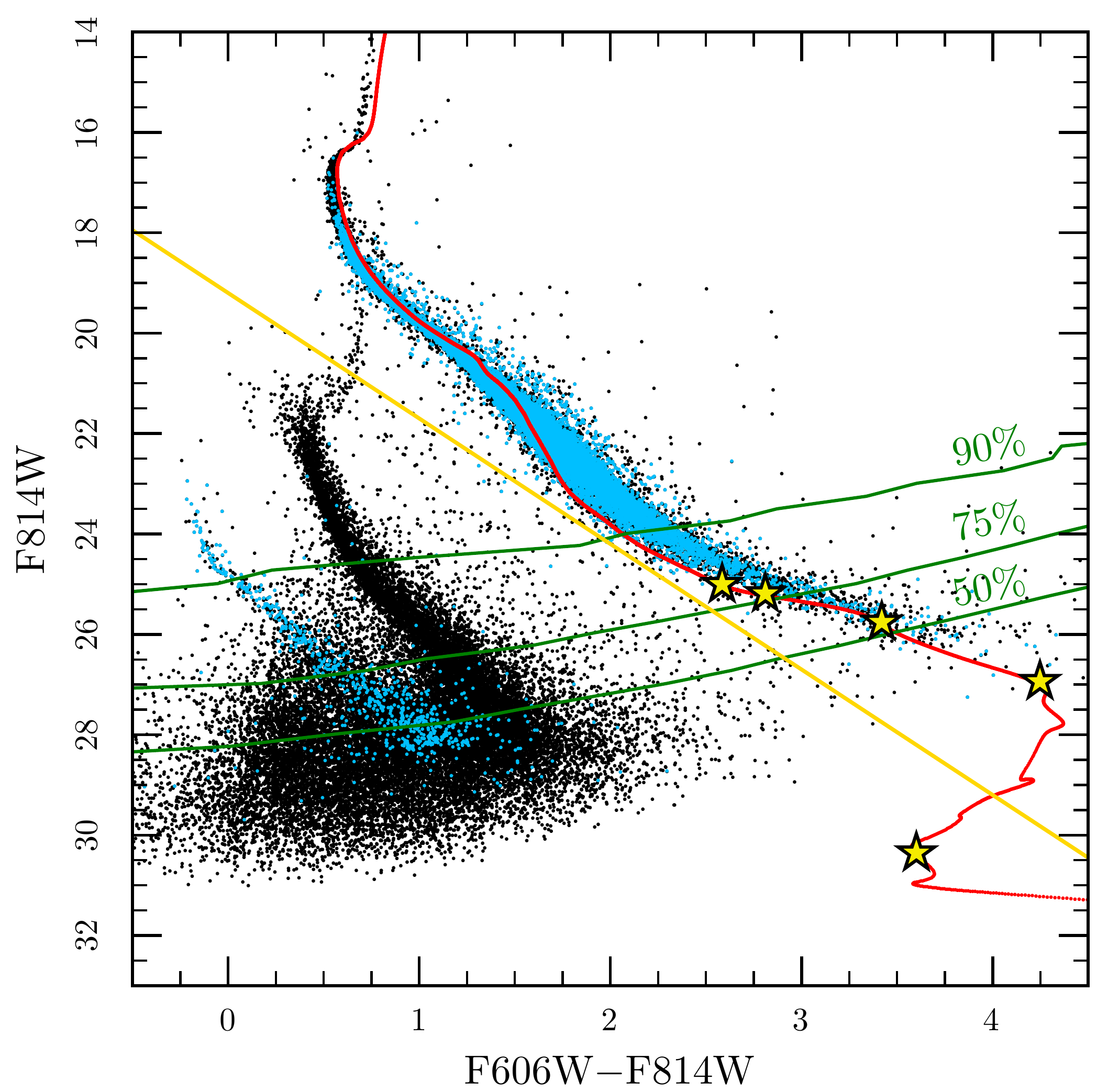}
\caption{CMD in Hubble's filters F814W and F606W for the ACS field described in \citep{1538-3881-143-1-11}. F814W is centered around 800~nm and F606W is centered around 600~nm (both with FWHM of about 150~nm). Light blue dots: stars known to be part of the cluster from observed proper motion; black dots: other stars and galaxies; red line: 11 Gyr isochrone built from our models. The three green lines trace the 90\%, 75\% and 50\% completeness rates (rates for detecting stars). We consider all stars regardless of their motion above the gold line to be main-sequence stars in the cluster.  This introduces a few stars from the SMC into our sample.  However, since we are interested in the low-mass end of the main sequence, this slight contamination is not important. The 5 yellow stars indicate the position of 5 stars evenly spaced in mass between, from bottom to top, 0.08 M$_\odot$ and 0.1 M$_\odot$. The kinks in the isochrone below 0.085 M$_\odot$ result from the formation of molecules in the cooler atmospheres.}
\label{fig:clean_cmd}
\end{figure}

With Hubble Space Telescope's deep observations of 47~Tucanae, we are already able to probe the faint end of the main sequence and white dwarf cooling sequence. As an example, Fig.~\ref{fig:clean_cmd} shows a CMD obtained from Hubble's Advanced Camera for Surveys (ACS) for the 47 Tucanae field described by \citet{1538-3881-143-1-11}.  The light blue points indicate stars that we know are moving across the sky like those in the cluster.  The black points are galaxies and stars that are either in the Small Magellanic Cloud, which lies in the background, or whose motions are too ill-determined to confirm cluster membership. On top, in red, we plot an 11 Gyr isochrone from our models for a distance of 4.7~kpc \citep{2012AJ....143...50W}. The transition region in this plot starts right where we have the faintest main sequence data points, around 27 magnitude in F814W.  Brown dwarfs, however, are very red, very faint and beyond even the reach of these ultra-deep observations with Hubble. Deep observations that reach the bottom of the main sequence are  available in Hubble's infrared bands \citep{2016ApJ...823...18C} but these measured suffer from crowding so it is difficult to determine definitively the number of stars as a function of mass from these infrared observations alone.  

With JWST's increased sensitivity and angular resolution it will be possible to observe the transition between low mass main sequence stars and brown dwarfs. In order to understand how this region will look in observations from JWST, we use the Phoenix atmospheres\footnote{\url{https://phoenix.ens-lyon.fr/Grids/BT-Settl/CIFIST2011\_2015/}} calculated by \citet{2015A&A...577A..42B} and \citet{2016Allard}. Again, the explanation of our method can be found in Appendx~\ref{sec:simJWST}. Models for atmospheres that cover the range of effective temperatures needed for the brown dwarf sequence are available in solar metallicity only. For future studies, it will be important to generate models for other metallicities and apply these both to the calculation of the spectra and the outer stellar boundary condition \citep{2015A&A...577A..42B}, to get a full understanding of how a different metallicity can change the picture.  

In this study, we picked the 47 Tucanae ACS field described by \citet{1538-3881-143-1-11} (see Fig~\ref{fig:clean_cmd}) as a suitable target for observing brown dwarfs.  Because the field is 6.7 arcminutes from the center of the cluster, it is not yet relaxed.  We calculate the relaxation time to be about twice the age of the Universe. Therefore, we do not expect that mass segregation has had time to expel low mass stars and substellar objects from the cluster.  The observations of this field with Hubble's Advanced Camera for Surveys comprise over 150~ks of exposure in both F606W and F814W.  This is one of the deepest observations of a globular cluster, and in fact, it probed almost to the very low mass end of the main sequence. Although we are going to look at brown dwarfs in the infrared bands with JWST, we opted for the visible observations of 47 Tucanae to perform our study. That is because Hubble's pixels are too big to achieve the needed sensitivity and resolution in the infrared to reliably count the faint stars at the lower end of the main sequence. This is exactly the problem that JWST will solve. The current absence of points below the end of the main sequence in Hubble's Wide Field Camera 3 infrared studies, such as \citet{2016ApJ...823...18C}, will be populated by detective brown dwarfs with JWST.

In order to predict how observations of brown dwarfs with JWST will look like in this field, we simulated 50,000 stars from our models, with a mass function inferred from the same field.  In Appendix~\ref{sec:simJWST} we explain the procedure and we show a CMD (Fig~\ref{fig:cmd}) for three different ages (9, 11 and 13 Gyr) in the JWST ultra wide filters F150W2 and F322W2. We have chosen these filters because they can be used for simultaneous observations and their large throughput is suitable for deep observations. As gap stars and brown dwarfs cool, they evolve along the sequence, so that the shape of the CMD does not change with age, contrary to what happens at the main-sequence turn-off \citep[see also][]{1997A&A...327.1054B}.  The transition region between the end of the main sequence and the beginning of the brown-dwarf sequence is located at an absolute magnitude of $\mathrm{F150W2} \sim 11.5$, where the density of stars becomes sparser. In these filters, it extends for almost three magnitudes in brightness and for more than one magnitude in color.

\section{Measuring the age of the cluster}
\label{sec:method}

\begin{figure}
\includegraphics[width=\columnwidth]{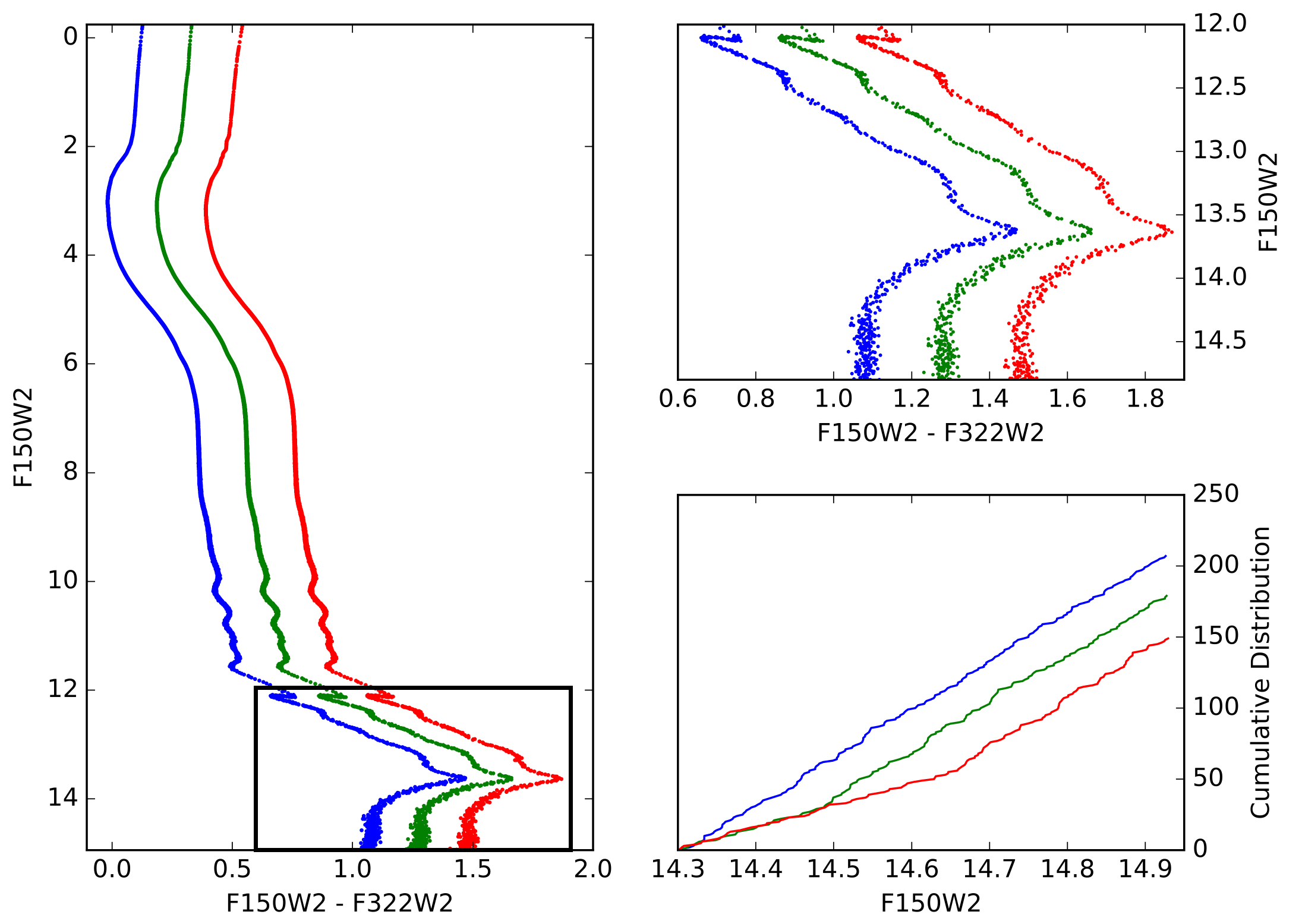}
\caption{Left: color-magnitude diagram for three different ages in the JWST ultra wide filters F150W2 and F322W2. F150W2 is centered around 1.7~$\mu$m (1.3~$\mu$m FWHM), and F322W2 is centered around 3.2~$\mu$m (2.6~$\mu$m FWHM). Blue: 11~Gyr; green: 12~Gyr; red: 13~Gyr. The 11-Gyr model was shifted 0.2 magnitudes to the left and the 13-model was shifted 0.2 magnitudes to the right so that the models could be distinguished. The black box highlights the region we focused on in the right panels. Upper right: zoom on the lower mass end of the transition zone.  The density of objects doubles at the top of the brown dwarf cooling sequence: magnitude 14.3 at 11~Gyr, of 14.5 at 12~Gyr and of 14.7 at 13~Gyr. Lower right: cumulative distribution for the three ages. Notice that the 12 and the 13 Gyr curves (green and red respectively) have the same slope up to 14.5 magnitude, then the 12 Gyr slope has a kink and it continues with the same slope as the 11 Gyr one (blue). The kink corresponds to the brightest end of the brown dwarf cooling sequence.}
\label{fig:3_cmd}
\end{figure}

In order to measure the age of the cluster, we focused on the brightest brown dwarfs, {\em i.e.} on the faint end of the transition region, shown in the zoomed CMD in Fig.~\ref{fig:3_cmd} (upper right panel).
To account for observational errors, we added a magnitude-dependent Gaussian noise to our simulated magnitudes, which corresponds to a signal to noise of 100 at magnitude 14.2 in F150W2 and at magnitude 13.1 in F322W2. In the plot, we shifted the 11-Gyr and the 13-Gyr CMDs 0.2 magnitudes blueward and redward respectively, to compare the differences between the ages.  In the box, we can see that the beginning of the brown dwarf cooling sequence gets fainter with age. The brightest brown dwarf (0.079 M$_\odot$) has a magnitude of 14.3 at 11~Gyr, of 14.5 at 12~Gyr and of 14.7 at 13~Gyr. Objects at fainter magnitudes (lower mass BDs) are more dense in the CMD.  Although it is subtle to see directly on the CMD in Fig.~\ref{fig:3_cmd}, it is clearly apparent in the cumulative distribution shown in the same figure (lower right panel).

In the plot, the start of the brown-dwarf sequence is evident as a steepening of the cumulative distribution, which corresponds to a larger number of objects per unit magnitude. We can see that the behavior in the range between 14.3 and 14.5 magnitude in F150W2 is similar for a 12 Gyr and a 13 Gyr cluster, while the number of objects for a younger cluster is much higher since the brown dwarf sequence has not cooled past this range yet. The number of stars in this range is 62 for the 11 Gyr old cluster and 30 for both the 12 and the 13 Gyr respectively (there is some randomness associated with these numbers which comes from the Gaussian noise). In the next range, between 14.5 and 14.7, the 12 Gyr cumulative distribution steepens, because that is where the top of the BD cooling sequence is at 12 Gyr. In this range, the number of objects is similar for the 11 and the 12 Gyr old clusters, 62 and 61 respectively, while it is less for the 13 Gyr, with only 32 objects. If the range chosen corresponds to the top of the BD cooling sequence, the value of the number of stars per unit magnitude doubles, and the top of the sequence gets fainter at about $0.2~\mathrm{mag~Gyr}^{-1}$ in F150W2. This can provide a precise estimate of the age of the cluster. For example, imagine that we observed a 12 Gyr old cluster with JWST. If we counted 30 stars in the 14.3-14.5 magnitude range, the Poisson probability of the cluster being 11 Gyr or younger would be $5\times 10^{-6}$. If we then counted the stars in the 14.5-14.7 range and we find a value around 60, the probability of the cluster being 13 Gyr or older is around $6\times10^{-6}$.  This means that we can determine the age of a cluster within one Gyr with a confidence of four sigma.

In order to pick the right range in absolute magnitude, the distance to the cluster should, in principle, be known.  However, there are some features in the CMD, due to the appearance of molecules in the stars' atmosphere, that do not depend on age.  An example is the hook in which stars in the transition region get bluer at $\mathrm{F150W2} \approx 12.1$ (Fig.~\ref{fig:3_cmd}).   The bright end of the transition region at $\mathrm{F150W2} \approx 11.5$ and the downturn in the CMD at $\mathrm{F150W2}\approx 7$ \citep{2016ApJ...823...18C} are also independent of age. If we can identify these features in the CMD, we can use them as a benchmark from which to measure the bright end of the brown dwarf sequence even if the distance to the cluster is unknown. Moreover, they can be useful to test the validity of the atmospheric and stellar models.

In our analysis we focused on 47 Tucanae, as it would be a perfect cluster to test the method. However, from the perspective of tracing the chemical evolution of our Galaxy, it would be interesting to observe and measure the age of other, more metal-poor clusters as well: ideally, some clusters with an intermediate metal content, like M5 and M13, and very metal poor clusters, like M30 and M92. In order to measure the age with the method presented in this work, the cluster in question has to be massive, with a long relaxation time and close enough that the brown dwarf sequence is not too faint.

\section{Conclusions}
\label{sec:conclusions}

A piece of information that is missing and that will affect our statistics is how the background will look in the brown dwarf region with JWST filters. The main unknown, which is also one of the science drivers of JWST, is what the first galaxies looked like and how common they were.  Unless they are very compact, these galaxies can be distinguished from the brown dwarfs by their images.  On the other hand, if they are compact and have similar colors and fluxes as the brown dwarfs, since what we want to observe is the number of brown dwarfs as a function of flux, we will have to estimate the density of galaxies in this region by observing a nearby blank field. In this way, we will be able to disentangle the two populations, but, in order to achieve the same statistics, we will have to observe a bigger field. Another possibility to obviate this problem is to observe the same field twice: in this way, it is possible to measure the proper motion of stars and then to identify the objects that belong to the cluster.

In any case, there are several reasons to believe that these very faint galaxies will not look like old brown dwarfs. The ACS field we focused on in this paper is contaminated by galaxies at a color of about 0.5 and 1.5 in F606W$-$F814W and at a magnitude of about 29 in F814W. The galaxies contained in the 2MASS Photometric Redshift catalog (2MPZ), described in \citet{0067-0049-210-1-9}, are much brighter (they are either closer or bigger), but their distribution peaks at about 1 as well in 2MPZ B$-$R and 0.5 in R$-$I (similar wavelengths to F606W and F814W). Their distribution in the 2MASS H filter (similar to JWST F150W2) and in the WISE W1 filter (similar to F322W2) peaks at about 0.7 in color and there are very few objects in the range where we expect to find brown dwarfs (between 1 and 1.8, see Fig.~\ref{fig:3_cmd}). This is reassuring as it suggests that galaxies will not contaminate our field in the color range that we are interested in. Furthermore, calculations performed by \citet{2011ApJ...740...13Z} for JWST filters seems to indicate that galaxies would approach the BDs color range only as they age for as much as a billion years, at which point they would be at a much fainter magnitude than brown dwarfs in Galactic clusters.

The method presented in this paper provides a new way to estimate the age of globular clusters. If employed simultaneously with other, independent, methods, like the main sequence turn-off fitting and white-dwarf cooling, it can put constraints on the age of the clusters and at the same time provide a test of our stellar models. In order for it to be the most effective though, we have to have a good handle on the stellar evolution and atmosphere models. Of particular importance is to develop a suite of realistic atmosphere models for different metallicities that extend to the range of effective temperature and surface gravity of brown dwarfs. The next step is then to employ these realistic values as boundary conditions for the stellar evolution models. If we have good control on all the pieces of the puzzle, we can simultaneously use all the different features in the CMD to test and improve our knowledge of stellar evolution and obtain an age and distance estimate to the cluster.

{\noindent \bf Acknowledgments}

The research discussed is based on NASA/ESA Hubble Space Telescope observations obtained at the Space Telescope Science Institute, which is operated by the Association of Universities for Research in Astronomy Inc. under NASA contract NAS5-26555. These observations are associated with proposal GO-11677. This research has made use of data obtained from the SuperCOSMOS Science Archive, prepared and hosted by the Wide Field Astronomy Unit, Institute for Astronomy, University of Edinburgh, which is funded by the UK Science and Technology Facilities Council. This work was supported by NASA/HST grants GO-11677, the Natural Sciences and Engineering Research Council of Canada, the Canadian Foundation for Innovation, the British Columbia Knowledge Development Fund.  It has made used of the NASA ADS, arXiv.org and the Mikulski Archive for Space Telescopes (MAST).

\bibliography{bd_evol_science}

\bibliographystyle{mnras}

\appendix

\section{MESA models}
\label{sec:mesamodels}

In order to simulate the observation of brown dwarfs we generated a suite of models with MESA stellar evolution code \citep{2011ApJS..192....3P} for different initial masses. Because our simulations focus on the field of
47~Tucanae described in \citet{1538-3881-143-1-11}, we use 47 Tucanae's metallicity, $Z=0.004$ \citep{2014MNRAS.437.3274V}.  The MESA instruction file or inlist we used is given in Tab.~\ref{tab:inlist}. Because the atmospheric boundary conditions for a convective star can change significantly its evolution, we experimented with the boundary conditions available in MESA. We found that the most realistic boundary conditions for our models were the \texttt{photosphere\_tables}. However, these tables do not go as low in effective temperature as needed for brown dwarfs. When the models get off the tables, they switch to the much simpler boundary conditions called \texttt{simple\_photosphere}, that simply fix the thermodynamic temperature to be the effective temperature at an optical depth of $2/3$. A realistic set of tables that includes low effective temperatures is \texttt{tau\_1\_tables}. These tables, however, are available for solar metallicity only. The models plotted in Fig. 1 in the main text were generated using these boundary conditions to show how stars evolve in the transition region between low mass main sequence stars and brown dwarfs.

\begin{table*}
\caption{MESA inlist for generating the stellar evolutionary models}
\label{tab:inlist}
\begin{multicols}{2}
{\scriptsize
\begin{verbatim}
&star_job

      show_log_description_at_start = .false.
      
      create_pre_main_sequence_model = .true.
      pre_ms_T_c = 2d5
      pre_ms_relax_num_steps = 1 

      kappa_file_prefix = 'gs98'

      change_initial_net = .true.      
      new_net_name = 'pp_extras.net'
      
      change_lnPgas_flag = .true.
      new_lnPgas_flag = .true.

/ ! end of star_job namelist

&controls

      initial_mass = 1.0
      initial_z = 0.004d0
      
      use_Type2_opacities = .true.
      Zbase = 0.004d0

      which_atm_option = 'photosphere_tables'
      
      dH_div_H_limit = 0.9d0
      okay_to_reduce_gradT_excess = .true.
            ! these are for calculation of 
             ! efficiency boosted gradT
            gradT_excess_f1 = 1d-4
            gradT_excess_f2 = 1d-2
            gradT_excess_age_fraction = 0.9d0
            
            ! turn on full all the time
            gradT_excess_lambda1 = 1.0
            gradT_excess_beta1 = 0.5
            gradT_excess_lambda2 = 1.0
            gradT_excess_beta2 = 0.5
            gradT_excess_dlambda = 1
            gradT_excess_dbeta = 1

      am_nu_visc_factor = 0
      am_D_mix_factor = 0.0333333333333333d0
      D_DSI_factor = 0
      D_SH_factor = 1
      D_SSI_factor = 1
      D_ES_factor = 1
      D_GSF_factor = 1
      D_ST_factor = 1
      
      varcontrol_target = 1d-3
      mesh_delta_coeff = 1.5

      photostep = 50
      profile_interval = 50
      history_interval = 1
      terminal_cnt = 10
      write_header_frequency = 10

      max_age = 15d9
      max_years_for_timestep = 1d9

      smooth_convective_bdy = .true.                  
      convective_bdy_weight = 1

      mixing_length_alpha = 2
      use_Ledoux_criterion = .true.
      alpha_semiconvection = 4d-2
      thermohaline_coeff = 2

      MLT_option = 'Henyey'

      overshoot_f_above_nonburn_core = 0.014
      overshoot_f_above_nonburn_shell = 0.014
      overshoot_f_below_nonburn_shell = 0.014
      overshoot_f_above_burn_h_core = 0.014
      overshoot_f_above_burn_h_shell = 0.014
      overshoot_f_below_burn_h_shell = 0.014
      overshoot_f_above_burn_he_core = 0.014
      overshoot_f_above_burn_he_shell = 0.014
      overshoot_f_below_burn_he_shell = 0.014
      overshoot_f_above_burn_z_core = 0.014
      overshoot_f_above_burn_z_shell = 0.014
      overshoot_f_below_burn_z_shell = 0.014

      overshoot_f0_above_nonburn_core = 0.004
      overshoot_f0_above_nonburn_shell = 0.004
      overshoot_f0_below_nonburn_shell = 0.004
      overshoot_f0_above_burn_h_core = 0.004
      overshoot_f0_above_burn_h_shell = 0.004
      overshoot_f0_below_burn_h_shell = 0.004
      overshoot_f0_above_burn_he_core = 0.004
      overshoot_f0_above_burn_he_shell = 0.004
      overshoot_f0_below_burn_he_shell = 0.004
      overshoot_f0_above_burn_z_core = 0.004
      overshoot_f0_above_burn_z_shell = 0.004
      overshoot_f0_below_burn_z_shell = 0.004
      
      RGB_wind_scheme = ''
      AGB_wind_scheme = ''
      RGB_to_AGB_wind_switch = 1d-4
      Reimers_wind_eta = 0.5d0  
      Blocker_wind_eta = 0.1d0

/ ! end of controls namelist

\end{verbatim}
}
\end{multicols}
\end{table*}

\begin{figure}
\includegraphics[width=\columnwidth,clip,trim=30 5 50 20]{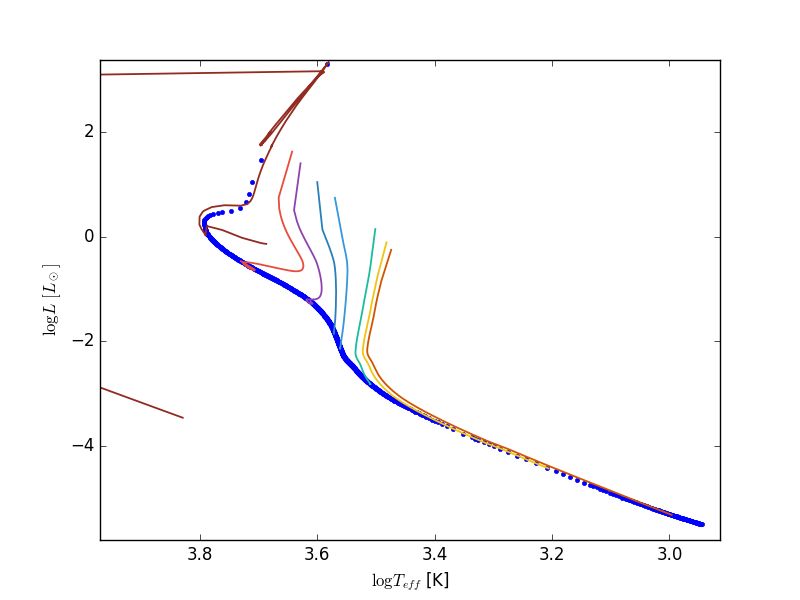}
\caption{The evolution on the HR diagram from pre-main sequence to 11 Gyr old of 8 stars with different masses is plotted in solid lines. The masses are from lowest to highest: 0.068, 0.08, 0.106, 0.207, 0.304, 0.502, 0.7, 0.974 M$_\odot$. Plotted in blue dots are 1000 stars, evenly spaced in mass between 0.06 and 0.88 M$_\odot$, all 11 Gyr old.}
\label{fig:models}
\end{figure}

Fig.~\ref{fig:models} shows our MESA models for different masses. The boundary conditions employed in this case are the \texttt{photosphere\_tables}. In the plot, all the models stop at 11 Gyr. From the interpolation of our models we generated isochrones, {\em i.e.} populations of stars with different masses but same age. In Fig.~\ref{fig:models}, plotted underneath the models (in blue), is an 11 Gyr isochrone of 1000 stars evenly spaced in mass, that we obtained from the interpolation of our models.  From the figure we can see the substantially different behavior between a low mass star (0.106 M$_\odot$) and a brown dwarf (0.068 M$_\odot$). Hydrogen burning keeps low mass stars bright while brown dwarfs cool down. A gap with very few objects is clearly visible in between. The stars in the gap are still in the pre-main sequence phase: nuclear luminosity contributes to their total luminosity, but the main contribution comes from gravitational contraction.

\section{Simulating observations with JWST}
\label{sec:simJWST}

MESA models provide information on the luminosity, effective temperature and surface gravity of the star at every stage of its life, but in order to simulate observations, we have to add atmospheres models on top to predict the fluxes in different filters. We use the Phoenix atmospheres\footnote{\url{https://phoenix.ens-lyon.fr/Grids/BT-Settl/CIFIST2011_2015/}} calculated by \citet{2015A&A...577A..42B} and \citet{2016Allard}. We integrate these spectra over the two HST filters in Fig.~1 in the main text and each filter planned for NIRCAM bands on JWST using the complete system throughput to calculate the broadband colors\footnote{\url{http://www.stsci.edu/jwst/instruments/nircam/instrumentdesign/filters}}.  We determined the absolute magnitudes in the Vega system, using the Alpha Lyrae spectrum built into PySynPhot as a standard.  Because 47 Tuc is only very mildly reddened, $E(B-V)=0.04$, we did not apply extinction to the NIRCAM fluxes, but we do apply extinction to the HST filters. This gives the magnitudes per unit area at a given effective temperature and surface gravity. Using these values from the stellar evolutionary models we interpolate through the atmosphere grid and calculate the absolute magnitude of the stars and brown dwarfs using the radius of the object also from the evolutionary model.

\begin{figure}
\includegraphics[width=\columnwidth,clip,trim=25 5 45 20]{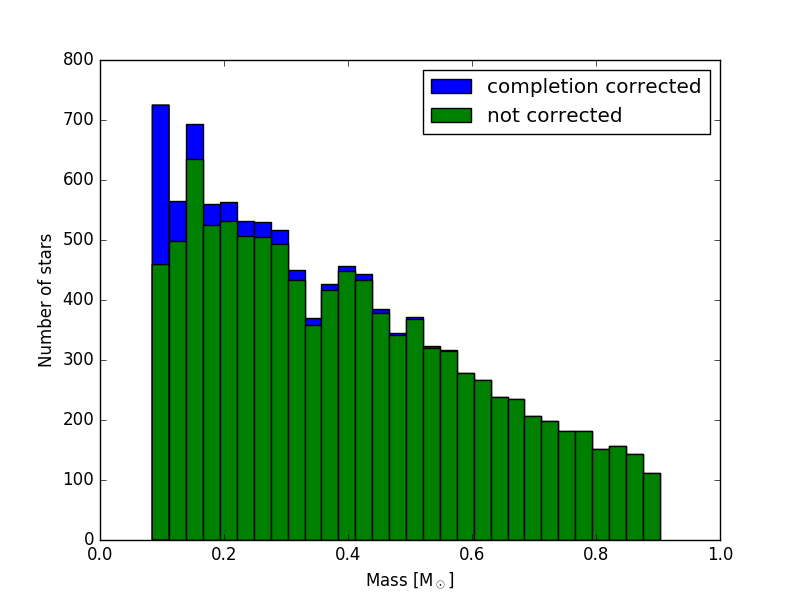}
\caption{Mass distribution obtained by comparing our isochrones with the ACS field described in \citep{1538-3881-143-1-11}}
\label{fig:mass_dist}
\end{figure}

In order to simulate JWST observations of the 47 Tucanae ACS field described by \citet{1538-3881-143-1-11}, we derived its mass function by comparing our models' magnitudes in F814W with the magnitudes observed in the field. We assumed a distance of 4.7~kpc \citep{2012AJ....143...50W}. The data we used is shown in Fig.~\ref{fig:clean_cmd}. The yellow solid line indicates the color-magnitude cut that we employ to get rid of the background galaxies, the white dwarfs and the stars in the Magellanic Cloud; the slope of the line is 2.5 and the x=0 value is 19.2. We assigned to each observed star in the field a mass corresponding to its magnitude in F814W, as inferred from the isochrone plotted in the same figure. While doing this, we restricted our analysis to the central part of the field.  The approximately circular region has a radius of 1.6~arcminutes, and the exposure is uniformly deep.  This is where we have a best estimate of the density of stars and their distribution in mass.  In spite of the depth of the exposures, we can still fail to find the faint stars within the field.  We estimate the rate of successfully finding stars, the completeness, with artificial star tests as outlined in \citet{1538-3881-143-1-11}, and we correct for the stars we miss (the completeness contours are shown in Fig.~\ref{fig:clean_cmd}). The mass distribution obtained is shown in Fig.~\ref{fig:mass_dist}. We used this mass distribution to simulate a field of 50,000 stars. To find 50,000 stars in 47 Tuc at this projected radius requires a survey of about $39~\mathrm{arcminutes}^2$ or four NIRCAM fields.

\begin{figure}
\includegraphics[width=\columnwidth,clip,trim=25 5 45 20]{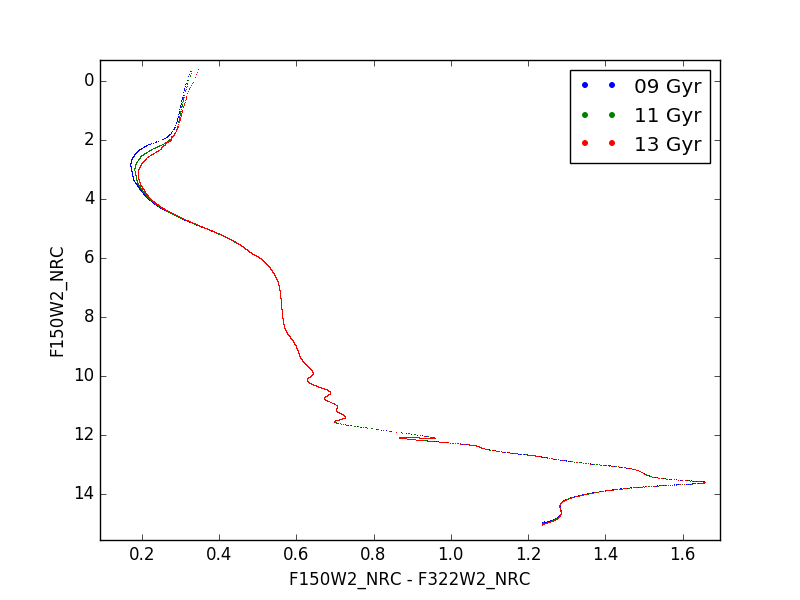}
\caption{Color magnitude diagram for three different ages in the JWST ultra wide filters F150W2 and F322W2.}
\label{fig:cmd}
\end{figure}

The final result of these simulations is the distribution of objects in the JWST ultra wide filters F150W2 and F322W2 as depicted in Fig.~\ref{fig:cmd} for three different ages (9, 11 and 13 Gyr). We have chosen these filters because they can be used for simultaneous observations and their large throughput is suitable for deep observations.

\label{lastpage}
\end{document}